# Supporting User Autonomy with Multimodal Fusion to Detect when a User Needs Assistance from a Social Robot

**Alex Reneau, Jason R. Wilson**

Northwestern University, Computer Science Department
alexreneau2021@u.northwestern.edu, jrw@fandm.edu

## Abstract

It is crucial for any assistive robot to prioritize the autonomy of the user. For a robot working in a task setting to effectively maintain a user's autonomy it must provide timely assistance and make accurate decisions. We use four independent high-precision, low-recall models, a mutual gaze model, task model, confirmatory gaze model, and a lexical model, that predict a user's need for assistance. Improving upon our four independent models, we used a sliding window method and a random forest classification algorithm to capture temporal dependencies and fuse the independent models with a late fusion approach. The late fusion approach strongly outperforms all four of the independent models providing a more wholesome approach with greater accuracy to better assist the user while maintaining their autonomy. These results can provide insight into the potential of including additional modalities and utilizing assistive robots in more task settings.

## Introduction

The autonomy of an individual is an essential priority of any assistive robot, especially when working with older adults (Lee & Riek 2018). In the context of task settings, an assistive robot needs to be able to provide reliable assistance to the user while allowing them to maintain a sense of pride and dignity. Particularly, medication sorting can be a stressful task for many older adults who have reduced executive functioning and increasingly need assistance. To help make the pill sorting task less stressful for older adults, four individual high-precision, low-recall models were built to enable a robot to help older adults sort their pills into a pill organizer by determining if they do or do not need help. If the user needs help, the robot provides key instructions, and if the user does not need help, the robot might present a motivational statement to encourage the user. Still, the primary goal of assistance is to provide reliable assistance while maintaining the user's autonomy.

In this paper, we present an approach to recognizing when a user needs assistance, which would allow the robot to provide timely assistance and support the user's autonomy. Starting with four high-precision, low-recall models that focused on specific cues in the interaction, we hypothesized that fusing the decisions of these models would yield greater performance because they collectively would capture the different ways in which a robot could detect when a user needs assistance. In finding an effective approach for fusion, we apply a sliding window method to better capture temporal dependencies and tested multiple learning algorithms. We compare the fusion model with the independent models, analyze the effect of the sliding window approach, and compared results of the learning algorithms. Overall, we demonstrate that the fusion model greatly outperforms the independent models and that a sliding window approach is effective in capturing the temporal dependencies in the data.

## Background

Many older adults face the stress of managing multiple medications, requiring them to employ a variety of strategies (MacLaughlin et al. 2005). One such strategy is using pill organizers, often ones with multiple days and multiple times in the day. Using pill organizers requires the person to sit down on a regular basis to reload the organizer. For some, this can be an arduous task, as numerous medications and vitamins can lead to many constraints that need to be satisfied. Additionally, many older adults have reduced executive functioning, increasing the need for assistance. Previous work has explored how a social robot may be able to assist a person in sorting their medications into a pill organizer (Wilson et. al. 2018).

One of the primary goals in the design of the robot was to maintain the autonomy of the user, thereby allowing a person to be involved in health-related decisions and contributing to a dignified quality of life. For the robot to support the autonomy of the person, it needed to be able to assess how and when to assist, so as to not provide too little or too much assistance. To guide the robot in making such decisions, it used a set of domain-specific rules. These rules were often fragile and could not be applied to other domains.

The current work looks to integrate a task model with a set of models to recognize social cues and enable a robot to learn how best to integrate the models such that the robot can recognize when the person needs assistance.

**Data collection**

Previously, data was collected from an experiment in which a social robot assisted people (N=19) in completing a medication sorting task. Each interaction was audio-video recorded, but three sessions had technical failures, resulting in 16 usable recordings (Wilson et. al. 2018). In each video, a user is tasked with organizing two medications onto a sorting grid. The robot monitors the task and user. When needed, the robot provides assistance in the form of utterances and gestures.

Each video was annotated to capture information about the user and the robot (Kurylo & Wilson 2019). Annotations consisted of gaze direction (a qualitative description of where the user was looking), user speech (what the user said), user gestures (whether the user was motioning with their hands), and robot speech (what the robot said). Additionally, task events representing each time the user placed or removed a pill were captured. Lastly, (Kurylo & Wilson 2019) recorded when the user in the video appears to need help, which is determined by a majority vote of three annotators. Table 1 shows how many events of each type were found in the 16 videos.

Table 1. Counts of each annotation type.

| Annotation type | Count |
|---|---|
| Gaze | 956 |
| User Speech | 145 |
| User Gesture | 220 |
| Robot Speech | 255 |
| Task | 402 |
| Help Needed | 872 |

## Models

**Independent Models**

A set of independent models, each working independently on the same data, were developed to recognize when a person needs assistance. The intent in developing these models was twofold: (1) development of the model can focus on particular features, (2) robot designs can integrate the appropriate models based on what features are available in the data. In this work, we integrate four independent models, three of which focus on social cues, and one that analyzes the user's progress in a task. Each of these models are briefly described below, and their performance is shown in Figure 1.

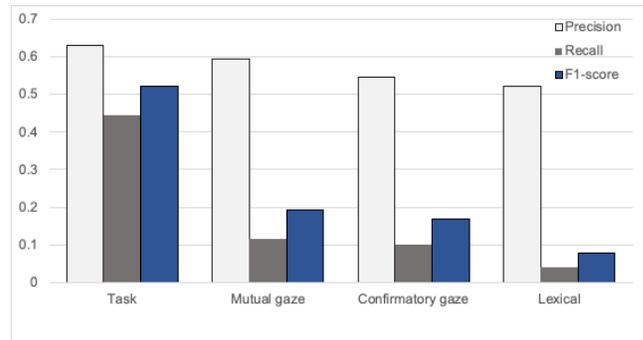

Figure 1. Each of the independent models have relatively high precision compared to their recall scores.

The three models that focus on social cues analyze eye gaze patterns and word choices and were developed based on patterns in human communication. For the models using eye gaze, initially four models were created based on patterns of eye gaze movement that were identified from the literature (Kurylo & Wilson 2019). These patterns consisted of mutual gaze, confirmatory gaze, gaze away, and goal reference. Of these models, mutual gaze and confirmatory gaze performed significantly better than the other models. The mutual gaze model produced a .59 precision and .12 recall for an F1-score of .19. The confirmatory gaze model had a precision of .55 and recall of .10 for an F1-score of .17. Only these two best models are integrated into the current work.

The language model uses keyword spotting to recognize likely indicators that the person is saying something that can be inferred to mean they need help. Without any semantic understanding or context, this model is not expected to be highly accurate. The precision of the model was .52 but the recall was only .04 and the resulting F1-score was .08.

A task-based model for recognizing when a person needs assistance utilizes changes in the number of steps to complete the task as an indicator of whether the person needs assistance (Wilson et. al. 2019). The model performs reasonably well (.63 precision, .44 recall, .52 F1), but it is clear that it is insufficient. A better model would integrate social cues (i.e., from eye gaze patterns and language) with progress in the task to have a holistic approach to recognizing when a person needs assistance.

**Fusion Model**

Our four independent models (task, mutual gaze, confirmatory gaze, lexical) have high precision and low recall, which gave us a strong indication that integrating these models would lead to increased recall and an overall improved performance. In this paper, we demonstrate that fusing these results does improve performance. The updated process for inferring whether a user needs help is shown in Figure 2. The annotated data is sent to each of the four individual models to be classified, and the decision outputs of each of the four models are packaged into feature vectors that are passed to a learning algorithm to be fused, providing us with our final decision output.

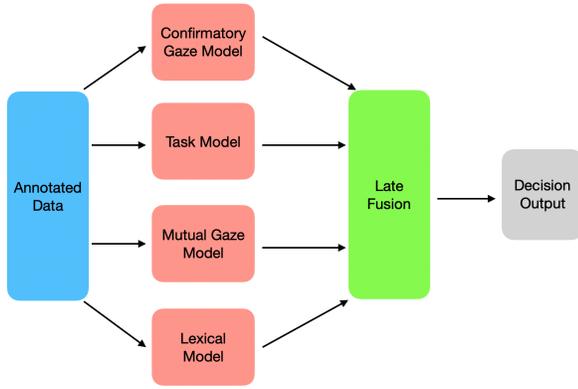

Figure 2. Annotations from the videos are processed by independent models. Decisions of the independent models are combined with a fusion model.

Since we are working with the outputs of other models, we decided to implement a late (or decision-level) fusion approach (Poria et. al. 2017). Late fusion is not the only fusion approach that is available. Early (or feature-level) fusion takes the features of each base-level model and compiles them into a feature vector that is passed along to a classifier to make the final decision (Atrey et. al. 2010). Still, an early fusion approach does not benefit our work because the independent models can be refined and adapted for generalizability, which will enable us to learn how the indicators detected by each model interact during a task for a given user, and it will allow for more reuse, minimizing what needs to be learned.

**Sliding Window Method**

A sliding window method is a machine learning technique that takes advantage of temporal data by using previous decisions or features to affect the current decision. A sliding window method was used in action recognition for industrial applications to help find weak spatio-temporal descriptors to reduce computational effort and dimensionality (Akkaladevi & Heindl 2015). Likewise, we use a sliding window method to capture temporal dependencies, but we do not prune weak spatio-temporal descriptors. Furthermore, a sliding window method is also used for gesture-based recognition in warehouses to provide warehouse robots with more context to better discern gestural instructions from warehouse workers (Neto et. al. 2019). Similarly, in our case we implemented a sliding window method to provide more context to better discern a person's need for assistance. Our sliding window method assists to capture the temporal dependencies in our data and enables us to use standard machine learning classification algorithms (Dietterich 2002).

There are many different implementations of the sliding window method found in other domains. For example, a sliding window method was used in stock forecasting to capture recent stock information to help predict next-day market closing prices (Hota et. al. 2017). However, our approach differs because our employed system will not have access to ground truth values. Since we do not know the ground truth values of previous interactions, we have to rely primarily on previous features from previous events. Our version of the sliding window method does not affect the target values associated with each feature vector. It only expands the range of every feature vector to account for previous information. Additionally, sliding window methods have been used in data mining to better capture temporal relationships across transactions (Li & Lee 2009). Our sliding window method is implemented in a similar manner such that we treat each feature vector like a transaction, but our data is structurally different from transactional data, so our output after the sliding window method is different.

Our implementation of the sliding window method consists of concatenating the specified amount of previous feature vectors to the current feature vector to preprocess our data.

$$\hat{X}_t = X_t, X_{t-1}, X_{t-2}, \ldots, X_{t-(s-1)}$$

$X$ represents all of the feature vectors, $t$ is the current event, and $s$ is the window size parameter that determines the number of windows per feature vector or how many previous feature vectors the current feature vector will include. $\hat{X}$ represents our new set of feature vectors. Each feature vector, without using history, has the size of 1 window because the length of each window is equal to the length of the feature vectors. $X_t$ is a feature vector in $X$. For each $X_t$, we append the $s-1$ previous feature vectors to $X_t$, which gives us $\hat{X}_t$. After applying our sliding window method to every $X_t$, we are left with the modified feature vectors, $\hat{X}$.

**Learning Algorithms**

Given our limited amount of data, we focus on evaluating learning algorithms that are data efficient. We consider five different algorithms: support vector machine (SVM), logistic regression, decision tree, naïve Bayes, and random forest (ensemble model). We use a Gaussian model for the naïve Bayes. After empirical testing, we selected a Radial Basis Function (RBF) kernel for the SVM. For the random forest, we use 1600 estimators, a max depth of 20, bootstrap samples, and included a random state.

## Experimental Procedure

To determine if the fusion model is better than the independent models, we conduct a series of experiments using the results of the independent models. In the experiments, we vary the size of the sliding window and which learning algorithm is used. Each experiment consists of three steps: preprocess the data to build windowed feature vectors, shuffle the windowed feature vectors, and performed a cross validation.

**Preprocessing data**

We load data for 16 videos of a user interaction with a robot. The data consists of results from each of the independent models and a corresponding target value. Each model outputs a continuous value $0.0 - 1.0$, representing the model's estimate of how much need for assistance the user has (0.0 represents no help needed, 1.0 means help is definitely needed). Target values are Boolean, representing whether or not annotators judged the user to be needing help at that moment.

Data corresponding with each video is loaded and preprocessed independently. For each video $i$, we construct the list of initial feature vectors $X_i$. To create a windowed feature vector $\hat{X}_{i,t}$, where t is the ordinal of the event within video $i$, we append $s$ previous feature vectors, where $s$ is the size of the window. More formally, we represent each windowed feature vector as follows:

$$\hat{X}_{i,t} = X_{i,t}, X_{i,t-1}, X_{i,t-2}, \ldots, X_{i,t-(s-1)}.$$

In the case that $t < s$, we pad $\hat{X}_{i,t}$ with a zero vector to account for missing windows. Our final set of features is $\hat{X}$.

**Shuffling Data**

Once the data is preprocessed and each windowed feature vector is constructed, we randomly shuffle the data to more evenly distribute true/false data points. Randomly shuffling the data does not disrupt temporal dependencies because the sliding window has already been applied. We shuffle the data to more uniformly distribute data before conducting our cross-validation. Without shuffling, results are skewed by the dominance of a few videos.

**Validation**

A 10-fold cross-validation is used for training and testing. During the training phase, we account for a slight skew in the data by balancing it to have an equal amount of true and false labeled data. For each fold, the balancing works as follows: Our new balanced training set will be {A, B}. (1) Set A is composed of all the true data points in the training set for the fold, and A has a size of $n$. (2) Set B is composed of $n$ randomly sampled false data points from the original training set for the fold. We did not balance the test set because skewed data is more prevalent in a testing environment with extended periods of the user not needing help.

During testing, we collect the confusion matrix for each windows size and algorithm combination. We then repeat the shuffling and validation 49 more times (a total of 50) to account for the randomization incurred by shuffling. We aggregate the results of each of the 50 iterations and calculate our key statistics: F1-score, ROC (Receiver Operating Characteristic), and PR (Precision-Recall).

After 50 iterations were complete, the whole process is repeated, starting with preprocessing the data, to build new windowed feature vectors for a different window size, which was varied 1-50. Overall, we ran 12,500 experiments (50 window sizes, 5 algorithms, 50 iterations).

**Results**

After running experiments that varied the window size and algorithms, we found that the window size plays a crucial role because it enhances the ability of all the algorithms to better capture temporal dependencies, which directly leads to improved results. The window sizes that most impact the rate of change of performance fall within the range 1-10. Our best performing algorithm was the random forest with the window size parameter set to 47. The random forest strongly outperforms all of our other optimized algorithms and is significantly better than our baseline model. The tree algorithms (random forest and decision tree) show more consistent performance growth as the window size is

Table 2. F1 scores of models for various window sizes

| | Window Size | | | | | | | | |
|---|---|---|---|---|---|---|---|---|---|
| | 1 | 10 | 20 | 30 | 40 | 41 | 47 | 48 | 50 |
| SVM | .62 | .66 | .68 | **.70** | .69 | .69 | .70 | .69 | .69 |
| Logistic Regression | .62 | .68 | .68 | .70 | .70 | .70 | .70 | **.70** | .70 |
| Decision Tree | .62 | .69 | .71 | .72 | .73 | .73 | **.74** | .74 | .74 |
| Naïve Bayes | .62 | .67 | .68 | .72 | .73 | **.73** | .73 | .73 | .73 |
| Random Forest | .63 | .73 | .77 | .79 | .81 | .81 | **.82** | .82 | .82 |
| Average all models | .63 | .69 | .70 | .72 | .73 | .73 | **.74** | .74 | .73 |

*Note: Window size with greatest F1 score is bolded.*

increased compared to the non-tree algorithms (SVM, logistic regression, naïve Bayes).

**Comparison with Independent Models**

Our results show that a fusion approach considerably improves upon the performance of the independent models. The best F1-score of the independent models was .52, for the task-based model. To validate our fusion model, it needs to have a better F1-score. Our results, displayed in Table 2, show that all variants of the fusion model outperform the best of the independent models. This result validates our hypothesis that integrating the high-precision, low-recall models yields a better performing model.

**Impact of Window Size on Performance**

To examine the effects of the window size on performance and compare algorithms for each window size, we will use F1-score as our metric of comparison. We included the average of the algorithms in Figure 3 and Table 2 to focus further on the relationship between the window size and performance.

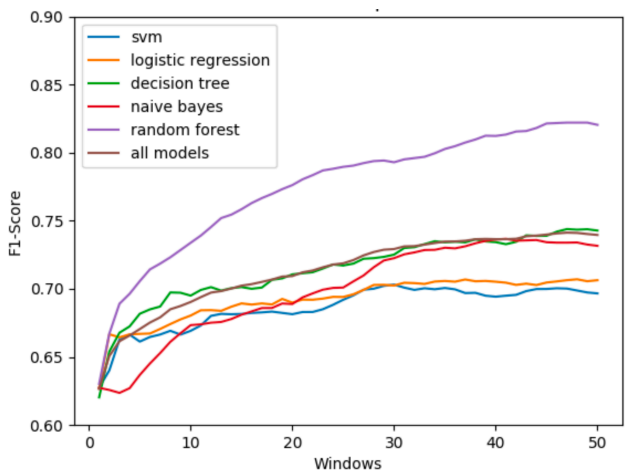

Figure 3. Made with 50 windows. For each window, the corresponding F1-score is the weighted average of 50 F1-scores to reduce variability from randomization. The 'all models' curve represents the average of all algorithms for each window.

From our findings, it is evident that the window size positively impacts performance regardless of any particular algorithm. As seen in Figure 3 and Table 2 by looking at the average of all algorithms, the F1-score increases as the window size is increased. From this, we can see that using more windows is effective in capturing the temporal dependencies in the data.

Window sizes 1-10 have the biggest impact on performance. All of the algorithms show improvement within that range. Looking at Table 2, if we take the difference of the F1-scores of the average of all algorithms for window sizes 1 and 10, it becomes evident that on average the F1-score increases by .07. The random forest algorithm shows the greatest improvement in the 1-10 window range increasing from .63 F1-score when the window size is 1 to .73 F1-score when the window size is 10. Conversely, the SVM shows the smallest improvement in the 1-10 window range increasing from .62 F1-score when the window size is 1 to .66 F1-score when the window size is 10. None of the algorithms show strong improvement when the window size parameter is greater than 48 and the results begin to plateau (Figure 3).

All of the algorithms are affected differently by the window size parameter, but the tree algorithms appear to show more consistent growth in performance as the window size is increased, and they plateau at larger window sizes. The tree algorithms peak at window size 47 with an F1-score of .74 for the decision tree and .82 for the random forest. Both tree algorithms begin to plateau after window size 47. We take the difference of the F1-scores for SVM, logistic regression, and naïve Bayes at window sizes 1 and 10 to see how much the scores increase from window sizes 1 to 10. SVM increases by .04 points, logistic regression increases by .06 points, and naïve Bayes increases by .05 points, which shows that all of the non-tree algorithms spike in performance until window size 10. Although, for windows 10-30, SVM and logistic regression show more steady performance, which we can see by taking the positive difference of the F1-scores of SVM and logistic regression at window sizes 10 and 30 (Table 2). SVM F1-score increases by .04 and logistic regression F1-score increases by .02 from windows 10-30. On the other hand, naïve Bayes experiences a spike in performance from window sizes 20-30. Referring to Table 2, the spike in performance becomes clearer by taking the difference of the F1-scores of naïve Bayes at window sizes 20 and 30, which shows naïve Bayes' F1-score increases by .04 from windows 20-30. Overall, the tree algorithms appear to have a more consistent relationship with the window size being changed.

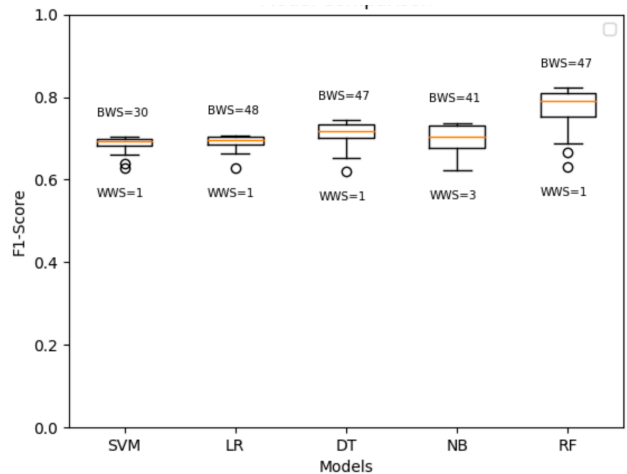

Figure 4. For each window, the corresponding F1-score is the weighted average of 50 F1-scores to reduce variability from randomization. BWS=Best Window Size. WWS=Worst Window Size. LR=logistic regression. DT=decision tree. NB=naïve Bayes. RF=random forest.

**Algorithm Comparison According to Window Size**

To compare the best version of each algorithm, we compare the distribution of F1-scores for each algorithm (see Figure 4) to identify the best (window size, algorithm) pairs. Our best performing (window size, algorithm) pair is (47, random forest) achieving 0.82 F1-score, our second best (47, decision tree) with 0.74 F1-score, our third best (41, naïve Bayes) with 0.73 F1-score, our fourth best (48, logistic regression) with 0.7 F1-score, and our fifth best (30, SVM) with 0.7 F1-score.

We use the F1-score for each windows size and algorithm combination (see Figure 3 and Table 2) to help determine the best algorithm and provide a progressive comparison for the algorithms as the window size is increased. Looking at Figure 3, all of the algorithms have relatively comparable scores when the window size is set to 1. The random forest begins greatly outperforming the other algorithms when the window size parameter is set to 10 with an F1-score of 0.73. When the window size is set to 10, the random forest's closest rival is the decision tree, which has an F1-score of 0.69. Although, in the window size range of 20-30, the naïve Bayes algorithm makes a jump to become more comparable with the decision tree algorithm. When the window size is set to 30, naïve Bayes has an F1-score of 0.72 and decision tree also has an F1-score of 0.72. Still, as the window size reaches 47, decision tree achieves an F1-score of 0.74 and naïve Bayes gets an F1-score of 0.73, which rules that decision tree is preferable to naïve Bayes. For all window sizes 1-50, SVM and logistic regression do not achieve F1-scores greater than 0.7. Therefore, SVM and logistic regression are inferior compared to decision tree and naïve Bayes, and we can conclude that random forest is clearly our best algorithm.

Table 3. Area Under the Curve (AUC) for ROC and PR curves

|  | ROC-AUC | PRC-AUC |
| --- | --- | --- |
| SVM | .74 | .68 |
| Logistic regression | .74 | .69 |
| Decision tree | .78 | .73 |
| Naïve Bayes | .75 | .65 |
| Random forest | .80 | .75 |
| Random (baseline) | .50 | .45 |

**General Algorithm Comparison**

For a more general comparison of the algorithms over all windows, we compare the Receiver Operating Characteristic (ROC) and Precision Recall (PR) curves (Figure 5). Using the ROC and PR curves, we calculated each of the corresponding AUC scores for each algorithm (Table 3). We also included a baseline model for comparison. The baseline model is a random model that randomly predicts true or false with equal probability. It achieves a ROC – AUC of 0.5 and a PR – AUC of 0.45.

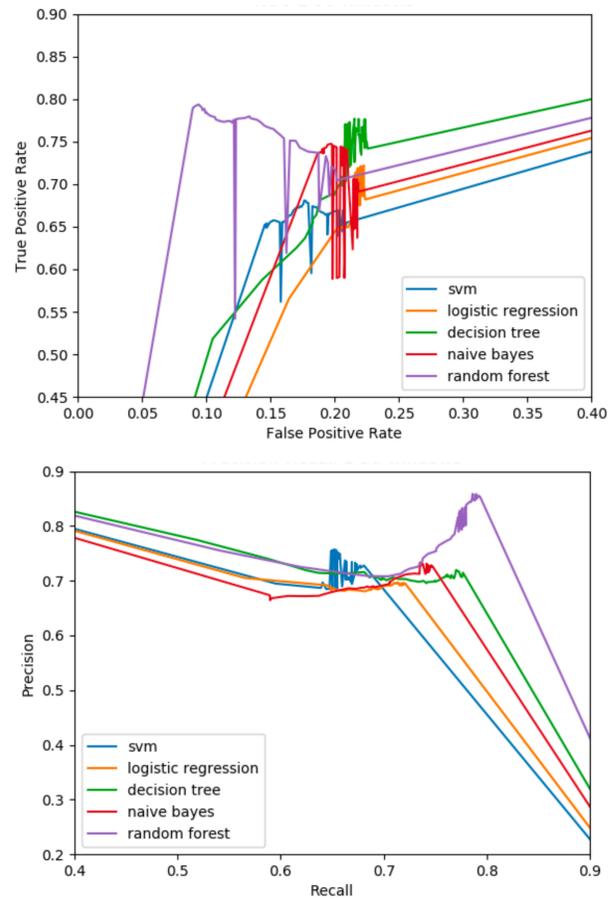

Figure 5. Receiver Operation Characteristic (ROC, above) and Precision-Recall (PR, below) curves. Made by varying window size 1-50.

Looking at Table 3, over all windows tested, our best performing algorithm is the random forest with a ROC – AUC of 0.8 and a PR – AUC of 0.75. The decision tree algorithm is the second overall best algorithm with a ROC – AUC of 0.78 and a PR – AUC of 0.73. There is then a small gap between the tree algorithms and the rest of the algorithms, where the ROC-AUC is between .74 and .75 and the ROC-AUC ranges from .65 to .69. However, all of these algorithms are effective because even our worst algorithm (naïve Bayes) greatly outperforms our baseline model.

Notably, the tree algorithms are more consistent over all window sizes and are the most reliable algorithms. If we take the positive difference of the ROC – AUC scores for the decision tree algorithm and naïve Bayes, we can see that the tree algorithms outperform the non-tree algorithms by at least 0.03 for ROC – AUC. If we take the difference of the PR – AUC scores for the decision tree algorithm and naïve Bayes, we can see that the tree algorithms outperform the non-tree algorithms by at least 0.04 for PR – AUC (Table 3). Conclusively, the tree algorithms are better than the non-tree algorithms.

## Discussion

The basis of our work is to create a better human-robot-interaction by supporting user autonomy in a task assisted by a social robot. There are many challenges in such a scenario, including accurately detecting the appropriate times for a robot to provide assistance. Additionally, interactions with the robot may be limited, thus constraining the amount of data that can be collected. Our approach demonstrates that we can integrate different models focused on specific cues to create a reliable multimodal fusion model that can accurately detect when a user needs assistance and does not require an abundance of training data.

### Great Improvement over Independent Models

Our results prove that a fusion approach considerably improves upon the performance of our four independent models. In addition to the worst fusion model outperforming the best individual model, the degree to which it outperforms exceeded our expectations. Clearly, multiple models were necessary to capture the different ways for a robot to recognize when a user needs assistance. While integrating the independent models was effective, to further improve upon the performance of the fusion model we need to consider the temporal dependencies in the data.

### Sliding Window Captures Temporal Dependencies

One of the most crucial factors to our work has been the window size parameter, demonstrated by its large impact on performance. We believe that the window size improves performance because windowed feature vectors account for more context than our original feature vectors. The features we worked with were the decisions of four independent models: mutual gaze, confirmatory gaze, task, and a lexical model. The mutual gaze, confirmatory gaze, and task models include a relative amount of context to make each decision. The lexical model does not include any context. Even though these models include some context, it is not sufficient enough to fully capture essential dependencies that our sliding window method can account for. Similarly, a sliding window method was used to detect the gestures of warehouse workers to inform and direct warehouse robots to better capture spatio-temporal variability even though it utilizes a LSTM deep net, which already captures temporal dependencies (Neto et. al. 2019).

To assess the impact of the window size, we experimented with a range of sizes and found that window sizes 1-10 induced a notable spike in performance. This demonstrates how critical it is to incorporate at least a minimal amount of context into these decisions. Similarly, when using a sliding window to capture temporal dependencies for human-action recognition for industrial purposes there is a spike in performance during window sizes 5-15 (Akkaladevi & Heindl 2015). Not including sufficient temporal context is then likely to lead to poor performance, such as a directions-giving robot that tracked user engagement (Bohus, Saw & Horvitz 2014). They concluded that more context is needed to account for temporary changes in the scene.

While smaller window sizes were most impactful, larger sizes do provide better performance. However, there is a cost in utilizing larger window sizes. As with most optimization problems, there is a give and take relationship when it comes to performance and runtime. It is very clear that as we increase the window size, the amount of time to train the fusion model increases. It is important that we strike a balance between performance and time, as there may not always be sufficient time to train a model in a deployed robot. Additionally, if we want the model to update and adapt to a particular user, the learning algorithm needs to be efficient.

A solution to this could be found in a human-gesture recognition study that made use of the sliding window method only during offline testing (Gu et. al. 2012). Another solution would be to optimize our sliding window method by using dynamic window sizes, which can provide accurate models while being more efficient with data and time (Laguna et. al. 2011). Similarly, applying a dynamic window size on our model will likely provide the accuracy and efficiency we need. In varying the window size, we expect that the optimal range will be between 10 and 47. Ten would provide efficiency and relatively good performance. Extending the window size beyond 47 would incur a greater cost, and it appears to have limited benefit. For our data, most of the algorithms exhibited a slight decline in performance when the window size exceeded 47. While it is not clear why the performance dips at this point, it is likely that this is related to the time it takes to perform steps in our task. With the average time between events being 1.5 seconds, a window size of 48 captures roughly 72 seconds of the task, which may coincide with the average amount of time it takes to sort a single type of medication. This suggests that information like the state of the task may be useful to mark how much context is needed to make an accurate inference.

### Data Characteristics Affect Learning Algorithms

We observed that the random forest and decision tree (i.e, tree-based algorithms) performed better than the other algorithms (SVM, logistic regression, naïve Bayes). While all of them perform relatively well with a minimal amount of data, the tree-based algorithms perform better, perhaps because they handle discrete and continuous data differently. Even though we are working with continuous data, the distribution of the values is heavily weighted at the extremes (i.e., most values are near 0 or 1). This makes the data more closely resemble a discrete structure, on which tree-based algorithms are likely to perform well. The SVM and logistic regression algorithms typically perform well with continuous input data, but the nearly binary nature of our data may contribute to these algorithms being less effective. Similarly, we use a Gaussian Naïve Bayes, which is designed for continuous data, but our data does not have a normal distribution, likely inhibiting the effectiveness of that approach.

There are a variety of algorithms we did not test that may have been useful because they inherently capture some temporal dependencies. For example, a Hidden Markov Model (HMM) requires a minimal amount of data to track changes in state. However, our task only has two states (needs help or not), making it difficult for an HMM to be effective without further decomposing each state. Also, HMMs do not scale well if the dimensionality is too high (Campbell et. al. 2019). Another problem with the HMM is that it has difficulty with obtaining temporal aligned sequences (Li et. al. 2017). Hence, the HMM does not appear to appropriately serve our problem compared to the other classification algorithms we chose.

Recurrent neural networks also capture temporal dependencies. For example, a LSTM deep net was used with skeleton data to help detect human activity (Li et. al. 2017). While, neural networks can be highly effective, they are prohibitively data and cost inefficient, making them unusable for our scenario.

**Late Fusion Over Early Fusion**

Our fusion model uses a late fusion approach to fuse our four independent models together. Other work has used both early and late fusion to monitor a robot's gaze control by fusing together audio-visual features, and it is notable that late fusion performed slightly better than early fusion (Lathuilière et al. 2019). While we could have experimented with an early fusion approach, it may not be a good approach for our setup. The late fusion allows us to leverage the knowledge already embedded in the independent models. This should contribute to requiring less data (and time) to train the fusion model (though we have not empirically validated this yet). We emphasize the cost to train the model because we envision that long term the fusion model will need to regularly update as it adapts to a user.

**Limitations**

One significant limitation of our work is that it pertains to a very specific domain. We have only focused on assisting people through a medication sorting task and cannot yet generalize our model to other task settings. Still, our findings yielded from the medication sorting task appear to be very promising and demonstrates great potential for being applicable in other task settings. It is also important to note that while we developed and tested within the medication sorting task, our work, particularly the independent models, is not necessarily tied to that task. For example, the eye gaze models look at general patterns of eye gaze, where these patterns are likely to exist in most assistive tasks but their prevalence and the meaning may vary between tasks. Learning the applicability of these models in relation to these other models is an ideal approach to using the fusion model we have demonstrated. To verify this, our future work will be focused on applying our models to other tasks, especially ones in which user autonomy and timely assistance by the robot are paramount.

Another limitation is that we are not able to work with large amounts of data since we are working with only a few interactions in a lab setting. We also do not expect we would be able to collect large amounts of data anytime soon, as there are few settings with a large amount of human-robot interactions. Adding to our data constraints is our plan to move towards having the robot adapt to an individual user. Adaptive behavior of the robot is a fundamental characteristic of an effective human-robot interaction (Mitsunaga et al. 2008), and we plan on implementing active learning approaches to allow our model to be updated based on ongoing interactions with the user. Others have shown that adaptive learning in can be effective and that users prefer an adaptive robot over a non-adaptive robot (Sekmen & Challa 2013). Knowing that we are always going to work with limited data and our plan to regularly update the model, we will continue to be limited to data and time efficient learning algorithms.

In order for us to move towards an adaptive learning approach, we need to collect feedback that is used to update the model. However, we do not have an effective way to collect this feedback, as asking the user would be inefficient and likely to be inaccurate. Similar to Mitsunaga et al., who used subconscious body signals to inform the robot how to adapt (2008), we need to develop a measure that can be used as a proxy. For example, signs of user frustration can indicate that the robot's help was unnecessary or poor.

## Conclusion

We tested multiple learning algorithms to use in a fusion model that integrates four independent models using a late fusion approach. Our experimentation found that a random forest algorithm effectively learns how to fuse the results of the models. Additionally, we found that a sliding window method captures the temporal dependencies in the data, significantly improving the performance of the fusion model. Our results show that a fusion model can help an assistive robot be more accurate in deciding when to assist in a medication sorting task. The timeliness of the robot's assistance would allow it to not be overbearing while also not leaving the user underinformed, thus empowering the user to author a successful completion of the task and thereby supporting the autonomy of the user.

# Acknowledgements

We would like to thank Ulyana Kurylo, Kevin Hou, and Roxy Wilcox for their contributions to the independent models and helpful comments.

# References

Akkaladevi, S. C., and Heindl, C. 2015. Action recognition for human robot interaction in industrial applications. *2015 IEEE International Conference on Computer Graphics, Vision and Information Security (CGVIS)* (pp. 94-99): IEEE.

Atrey, P. K.; Hossain, M. A.; El Saddik, A.; and Kankanhalli, M. S. 2010. Multimodal fusion for multimedia analysis: a survey. *Multimedia systems*, 16(6), 345-379: Springer.

Bohus, D.; Saw, C. W.; and Horvitz, E. 2014. Directions robot: in-the-wild experiences and lessons learned. *Proceedings of the 2014 international conference on Autonomous agents and multi-agent systems* (pp. 637-644).

Campbell, J.; Stepputtis, S.; and Amor, H. B. 2019. *Probabilistic Multimodal Modeling for Human-Robot Interaction Tasks*. arXiv preprint arXiv:1908.04955. Ithaca, NY: Cornell University Library.

Dietterich, T. G. 2002. Machine learning for sequential data: A review. *Joint IAPR international workshops on statistical techniques in pattern recognition (SPR) and structural and syntactic pattern recognition (SSPR)* (pp. 15-30): Springer, Berlin, Heidelberg.

Gu, Y.; Do, H.; Ou, Y.; and Sheng, W. 2012. Human gesture recognition through a kinect sensor. *2012 IEEE International Conference on Robotics and Biomimetics (ROBIO)* (pp. 1379-1384). IEEE.

Hota, H. S.; Handa, R.; and Shrivas, A. K. 2017. Time series data prediction using sliding window based rbf neural network. International *Journal of Computational Intelligence Research*, 13(5), 1145-1156: ripublication.

Kurylo, U., and Wilson, J. R. 2019. Using Human Eye Gaze Patterns as Indicators of Need for Assistance from a Socially Assistive Robot. *International Conference on Social Robotics* (pp. 200-210): Springer, Cham.

Laguna, J. O.; Olaya, A. G.; and Borrajo, D. 2011. A dynamic sliding window approach for activity recognition. *International Conference on User Modeling, Adaptation, and Personalization* (pp. 219-230): Springer, Berlin, Heidelberg.

Lathuilière, S.; Massé, B.; Mesejo, P.; and Horaud, R. 2019. Neural network based reinforcement learning for audio–visual gaze control in human–robot interaction. *Pattern Recognition Letters*, 118, 61-71: Science Direct.

Lee, H. R., and Riek, L. D. 2018. Reframing assistive robots to promote successful aging. *ACM Transactions on Human-Robot Interaction (THRI)*, 7(1), 1-23: ACM.

Li, H. F., and Lee, S. Y. 2009. Mining frequent itemsets over data streams using efficient window sliding techniques. *Expert systems with applications*, 36(2), 1466-1477: Science Direct.

Li, K.; Zhao, X.; Bian, J.; and Tan, M. 2017. Sequential learning for multimodal 3D human activity recognition with Long-Short Term Memory. *2017 IEEE International Conference on Mechatronics and Automation (ICMA)* (pp. 1556-1561): IEEE.

MacLaughlin E.J., Raehl C.L., Treadway A.K., Sterling T.L., Zoller D.P., Bond C.A. 2005. Assessing medication adherence in the elderly: which tools to use in clinical practice? *Drugs & Aging* 22(3):231–255

Mitsunaga, N., Smith, C., Kanda, T., Ishiguro, H., & Hagita, N. (2008). Adapting robot behavior for human-robot interaction. *IEEE Transactions on Robotics*, 24(4), 911–916.

Neto, P.; Simão, M.; Mendes, N.; and Safeea, M. 2019. Gesture-based human-robot interaction for human assistance in manufacturing. *The International Journal of Advanced Manufacturing Technology*, 101(1-4), 119-135: Springer.

Poria, S.; Cambria, E.; Bajpai, R.; and Hussain, A. 2017. A review of affective computing: From unimodal analysis to multimodal fusion. Information Fusion, 37, 98-125: *Science Direct*.

Sekmen, A., and Challa, P. 2013. Assessment of adaptive human–robot interactions. *Knowledge-based systems*, 42, 49-59: Science Direct.

Wilson, J. R.; Lee, N. Y.; Saechao, A.; Tickle-Degnen, L.; and Scheutz, M. 2018b. Supporting human autonomy in a robot-assisted medication sorting task. *International Journal of Social Robotics*, 10(5), 621-641: Springer.

Wilson, J. R.; Kim, S.; Kurylo, U.; Cummings, J.; and Tarneja, E. 2019. Developing Computational Models of Social Assistance to Guide Socially Assistive Robots. *Proceedings of the AI-HRI Symposium at AAAI-FSS 2019*.